\documentclass[a4paper]{article}
\usepackage[margin=3.5cm,nohead]{geometry} 
\usepackage{graphicx}
\title{The Multi-Core Era - Trends and Challenges}
\author{Peter Tr\"oger, Blekinge Institute of Technology}
\begin{document}
\maketitle
\begin{abstract}
Since the very beginning of hardware development, computer processors were invented with ever-increasing clock frequencies and sophisticated in-build
optimization strategies. Due to physical limitations, this 'free lunch' of
speedup has come to an end.

The following article gives a summary and bibliography for recent trends and challenges in CMP architectures. It discusses how 40 years of parallel computing research need to be considered in the upcoming multi-core era. We argue that future research must be driven from two sides - a better expression of hardware structures, and a domain-specific understanding of software parallelism.
\end{abstract}
\section{Introduction}
In 1995, Gregory F. Pfister formulated three ways of doing anything faster \cite{pfister}:

\begin{itemize}
	\item Work harder.
	\item Work smarter.
	\item Get help.
\end{itemize}

The original statement explained the benefits of parallel processing on cluster infrastructures. Instead of relying on the bounded scalability of a single machine ('work harder'), he proposed the usage of multiple machines working together as a cluster ('get help'). Within the last years, it turned out that Pfister's principles now also apply to a different area. 

Since the very beginning of hardware development, computer processors were invented with ever-increasing clock frequencies ('work harder') and sophisticated in-build optimization strategies ('work smarter'). Developers and industry got used to the fact that applications were simply getting faster by a simple exchange of hardware. Moore's law about the ever-increasing number of transistors per chip still applies, but brings no longer a better computational performance per default. The traditional way of increasing clock rates and higher packaging densities reached its physical limitation in cooling and power management \cite{olukotun:future:acmqueue:2005}.

In order to keep the promise of faster processing by hardware exchange, all CPU vendors now use the additional transistors for \emph{chip multi-processing (CMP)} architectures, which are currently named as \emph{multi-core} or \emph{many-core} processor design. A multi-core CPU combines multiple independent \emph{execution units} into one processor chip, in order to execute multiple instructions in a truly parallel fashion ('get help'). The cores of a CMP processor are sometimes also denoted as \emph{processing element} or \emph{computational engine}. According to Flynn's taxonomy \cite{flynn72}, the resulting systems are true multiple-instruction-multiple-data (MIMD) machines, able to process multiple threads of execution at the same time.  

Achieving processing speedup by using parallel execution units is obviously nothing new. The concept of the first \emph{multiprocessor computer} goes back to the 60's with initial architectures such as the ILLIAC IV \cite{illiac4}. The resulting challenges were also discussed to large extend, sometimes even 25 years ago \cite{cmpmyths}.

What is the difference ?
 
The true paradigm shift today is not the realization of CMP architectures, but their spreading in all computer markets. Embedded systems, mobile phones, desktop systems and server systems now include multiple cores out of the box. A parallel computer is no longer a dedicated hardware setup for special purposes. 

It is commodity.

While parallel computing of the past was only intended for a specific set of problem domains, it is now relevant for every scientific, industrial or private application running on a computer. Parallel computing now becomes visible for millions of industrial software developers in practice, who usually lack experience in the creation and handling of fine-grained parallel activities. For academia, the shift will influence both future teaching and research in computer science and software engineering. Many traditional areas such as computer hardware architecture, programming languages, design patterns, scheduling theory and parallel algorithms will gain more attention in the future, since the upcoming research challenges demand answers from these fields. In the following text, we want to provide a high-level overview about some of the identified issues in this area. We base the argumentation on one fundamental statement: 

\begin{quotation}
"40 years of parallel computing need to be considered." 
\end{quotation}

\section{Parallel Hardware}

\begin{figure}
\centering
\includegraphics[width=0.9\textwidth]{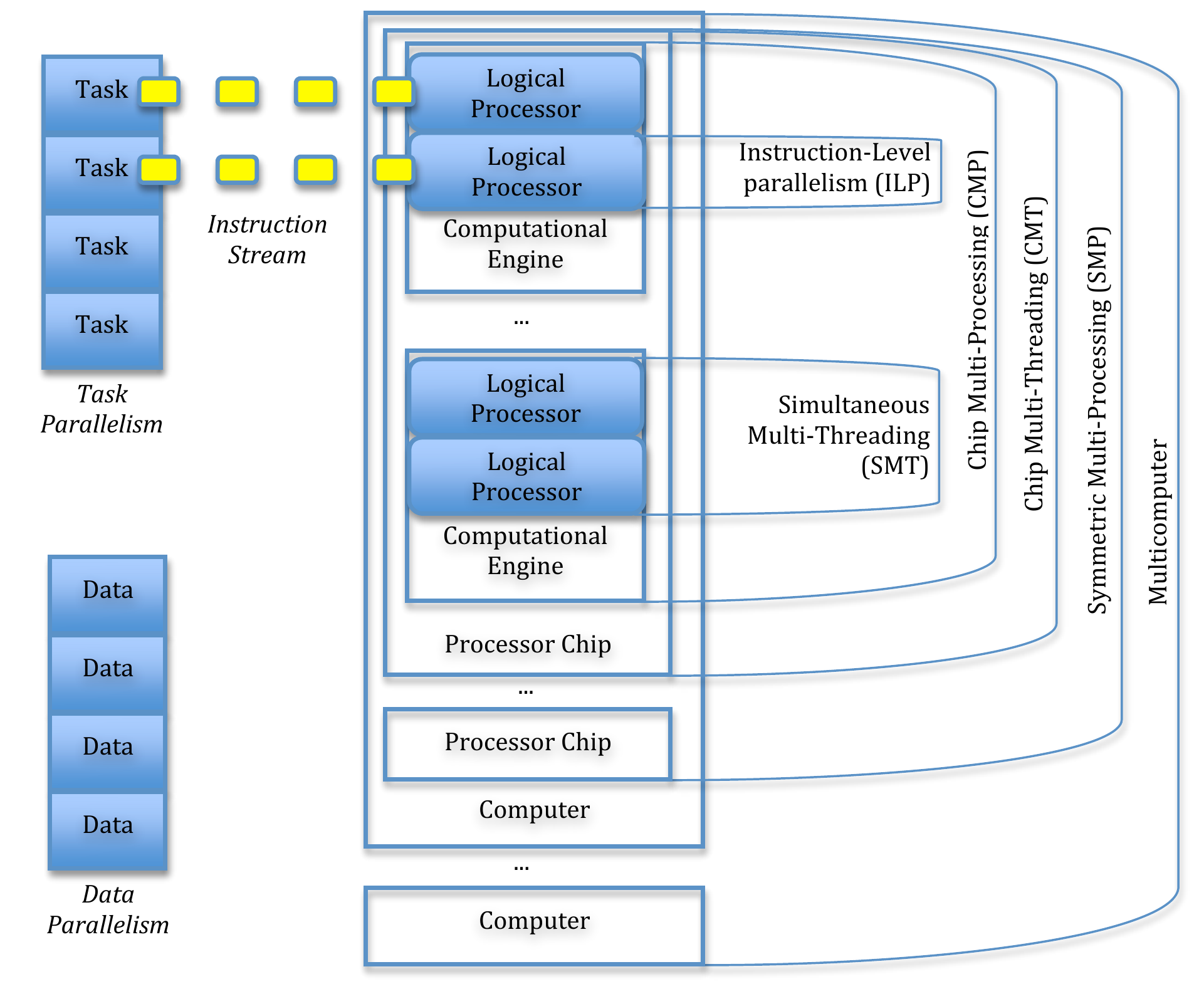}
\caption{Hardware parallelism hierarchy}
\label{fig:parhierarchy}
\end{figure}

The support for multiple parallel activities (\emph{multithreading}) in hardware is realized on different levels in todays computer systems (see Figure \ref{fig:parhierarchy}). The scheduling of parallel activities must now consider this given 'stack' of execution units. Even though modern operating systems are meanwhile aware of this effect \cite{hyperthreadingos}, they still cannot consider data dependencies between parallel activities in their scheduling decision. The application therefore has to support the operating system scheduler in the placement with its specific knowledge \cite{cmt}. 

On the lowest level, the execution unit itself can have a super-scalar architecture. A hardware-controlled parallel usage of execution unit components enables the execution of multiple processor instructions during one clock cycle. This approach of \emph{instruction-level parallelism (ILP)} is limited \cite{emer:single-vs-multi:ieee-micro:2007}, but widely realized approach in modern processor designs. 

Each execution unit can additionally support the concept of a logical processor, which allows a \emph{simultaneous multi-threading (SMT)} inside the processor pipeline \cite{224449,10.1109/40.621209}. This approach allows to hide data access latencies for some of the threads, by utilizing the execution unit for other tasks during a blocking period. This realizes a virtual sharing of one execution unit within a processor. SMT is better known under the marketing term \emph{hyperthreading} from Intel \cite{hyperthreading}. It maintains the architectural state of the execution unit (mostly CPU registers) separately per logical processor, in order to allow context switching in hardware. 

A set of execution units can be put together to form a \emph{chip multi-processing (CMP)} architecture \cite{mcdougall:sw-scaling:acmqueue:2005}. Most such processors contain separate L1 caches for each of the units, and a shared L2 cache for the whole processor chip. CMP forms the latest trend in processor design, even though earlier attempts already gained some experience with this approach \cite{mtproc}. 

The different ways of exploiting parallelism inside one processor chip are then summarized as \emph{chip multi-threading (CMT)} capabilities \cite{cmt}. The next higher level is \emph{symmetric multi-processing (SMP)} with multiple processors, and ultimativly the realization of a computer cluster as one \emph{multicomputer}. 

The widely promoted new era of multi-core systems basically focuses on the introduction of CMP features in standard desktop processors. Even though there is a high amount of daily news on recent hardware development, some common trends can be identified in this development. 

Most sources agree that the number of execution units per chip will be in the magnitude of thousands in the next 5-10 years. A recent report from Berkeley \cite{parcomplandscape} predicts CMP processors with thousands of parallel execution units as the mainstream hardware of the future. The "Intel Terascale Computing" initiative designed in 2008 an 80-core research prototype to investigate future challenges for the company. The MIT-originated TILE64 processor architecture today provides a grid of 64 low-speed (866 MHz) execution units on a chip, interconnected by a mesh network. In general, the trend with such \emph{homogeneus multi-core architectures} occurs to be a high number of less complex execution units, working together by some high-speed connection grid. 

The opposite approach is a \emph{heterogeneus multi-core architecture}. One example is the IBM Cell processor. It combines a single master processor element with a set of vector processors. The relevant property here is the heterogeneous set of execution units combined to one chip. It allows an prepared algorithm to use an optimized core for the computational task, while doing different activities in parallel. This kind of task-level parallelism is already common for modern personal computers with their dedicated chips for graphic processing and I/O, but now also reaches the CPU itself. Another well-known example is the zSeries mainframe concept \cite{zseries990}, combining different processing units on one \emph{multi chip module} with a shared L2 cache. Some of the units are dedicated as fault tolerance spares or diagnostic chip. This demands specialized compilers and an operating system prepared for the according processor platform. 

A still underestimated hardware problem is the connectivity of the execution units. The overall chip still must be connected to memory and I/O devices, which are an order of magnitude slower in their latency time \cite{mcore_bandwidth, mcore_rambandwidth}. Industry provides some solutions for this issue, which still need to be evaluated for higher numbers of execution units \cite{hypertransport}. 

The overall idea of many slow processing units connected by a high-speed bus is also not completely new. Computer hardware history shows the INMOS transputer concept \cite{may:95}, classical vector computers or massive parallel processing (MPP) installations. It is therefore necessary to consider the tremendous existing knowledge from these areas in the future.

A problem example in the hardware category is the contention of shared resources. With the increased introduction of parallelism in software layers, the processor faces a high number of threads with different memory access patterns and cache utilization profiles. Proposals by Intel suggest the introduction of a QoS-aware memory hierarchy \cite{qosmemhierarchy}, were the operating system prioritizes some threads in their cache and memory bandwidth usage. Other strategies still need to be investigated. This shows the general importance of a \emph{better software to hardware mapping}. Above the level of superscalar instruction processing, all parallelization coordination must now be done in software. This coordination must consider the given parallel hardware layers, without binding itself too much on one particular processor architecture.  It is therefore urgent to express modern hardware design in a more generic way, in order to develop appropriate scheduling and data management approaches for parallel applications. 

The history of high-performance computing provides the idea of an \emph{abstract MIMD machine model}, which is also applicable for modern CMT architectures. Such models support the run-time performance analysis for a given parallel algorithm, based on a very abstract understanding of a parallel execution environment. Famous examples are the LogP model \cite{culler93logp}, the bulk synchronous parallel (BSP) computer model \cite{bsp} and the parallel random access machine (PRAM) model \cite{pram}. 

Most of the existing models imply unlimited space or cycle time, and can even lead to contrary efficiency numbers for the same algorithm \cite{whatisscalability}. Other models such as the \emph{multicomputer} \cite{fosterparprog} focus only on distributed systems with multiple processors. It is therefore necessary to find a better abstraction of multi-core hardware architectures. This includes the consideration of timing effects reasoned by caches and mutual exclusion, which more and more turn out to be the performance-relevant factor. 

We therefore propose the development of a realistic and feasible \emph{multi-core machine model}, in continuation of the existing approaches. This model can provide the foundation for future research on parallelized algorithms and software design patterns.

\section{Parallel Software}

Multi-core systems are true parallel computers. The new trend therefore leads to the wider recognition of partially well-known and partially largely software problems. Industry and research already agree upon the fact that the existing \emph{programming paradigms and languages} as well as \emph{design patterns} do not align to modern processor design \cite{amdahlacm}.  This leads to the fact that the original hardware scalability problem is about to be replaced by a software parallelization problem. Challenges that are well known to the parallel computing community are now the problem for every software developer. 

This is nothing less than a paradigm shift in education, training, and daily practice of software development. Developers must get a basic understanding of parallel programming from the very beginning, treating a sequential program only as special case. This relates not only to tools and languages, but also to design patterns, algorithmic thinking, testing strategies and software architecture principles. The usage of multiple execution units has even influence on dependability issues \cite{mcoreisolation}, meaning that it also becomes a relevant aspect for research on fault-tolerant systems. 

\subsection*{Basic Principles}

The basic principles of parallel respectively concurrent programming are known for a long time. The mutual exclusion problem, deadlock, livelock and starvation are well-known questions for most computer science students and researchers. The challenge here is the consideration of such problems for the 'average' programmer. Due to the industrialization and wide-spread of software development, more and more developers do not have a solid scientific background in parallel computing. It is therefore necessary to assist these people in their development effort by an according testing mechanism, programming environment and high-level helper functionality. 

\begin{figure}
\centering
\includegraphics[width=0.9\textwidth]{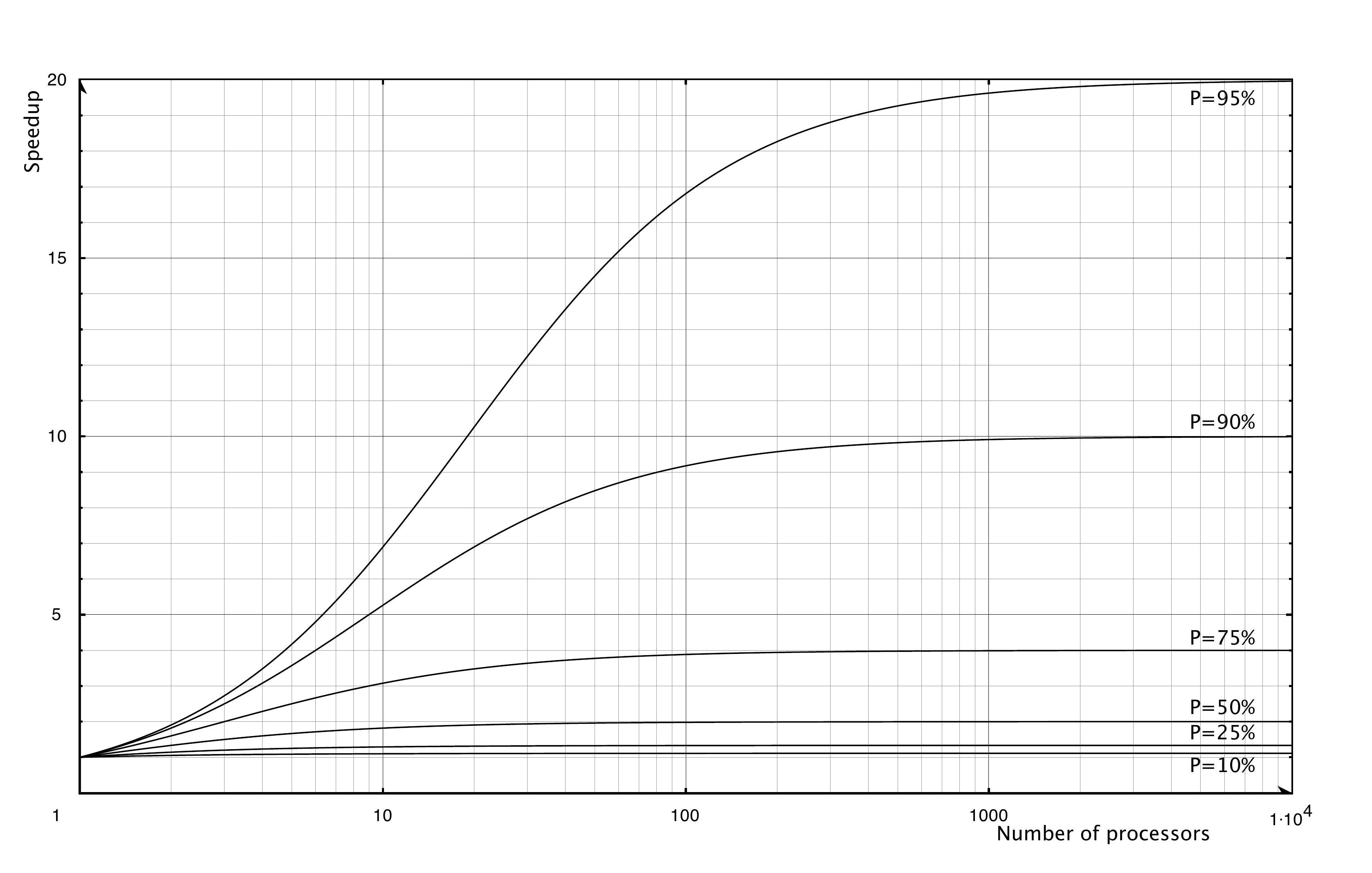}
\caption{Amdahl's law: Program with parallel portion P}
\label{fig:amdahl}
\end{figure}

There are widely accepted principles for the speedup achievable by parallelization. The two corner stones are Amdahls and Gustafsons laws. Amdahls law \cite{amdahl} puts an upper limit on the parallelization speedup achievable with a constant problem size. Figure \ref{fig:amdahl} shows thow speedup achievable by more execution units is always limited, even with a mostly parallelized application (P=95\%) . Gustafsons law \cite{gustafson} relaxes this upper bound by expressing that speedup can always be achieved by the shift to bigger problems. 

These two intuitive facts are still very valid in the modern multi-core era \cite{amdahlmulticore}, and therefore need to be considered in all related research. This demands a proper analysis of parallelization strategies for software. We propose that this kind of analysis can be for application domains in a whole, in order to derive generic parallelization principles and approaches. One example for this strategy is the traditional parallel database research, another one is the new area of parallelized XML processing \cite{head07, lu08}. Future work needs to identify more such application domains and should provide according parallelization strategies for them. This of course demands a heavy involvement of the according user communities, which is a new chance for computer science to be better connected to other sciences.  

\subsection*{Parallelization Support}

Parallelization support, already implemented by somebody else, can be provided through the compiler, the operating system or third-party libraries. Such support functionality is used by the application developer either implicitely (e.g. automated loop parallelization) or explicitely through new language constructs respectively a dedicated parallel language.

A compiler can try to formulate the assembler code in a way that multiple execution units are automatically used. In the perfect world, even the operating system developer would then be freed from the consideration of parallel execution strategies. Recent projects in this area \cite{polaris} still show that this remains a grand challenge of computer science. Since the parallelizing compiler has to determine possible side effects automatically, it becomes extremely hard to still generate valid code. Automatic parallelization mainly focuses on loops, since the identification of coarse-grained parallel chunks remains application-dependent. It is therefore a commonly accepted fact that parallelizing compilers can only provide limited help with speedup by parallelization. 

\begin{figure}
\centering
\includegraphics[width=0.5\textwidth]{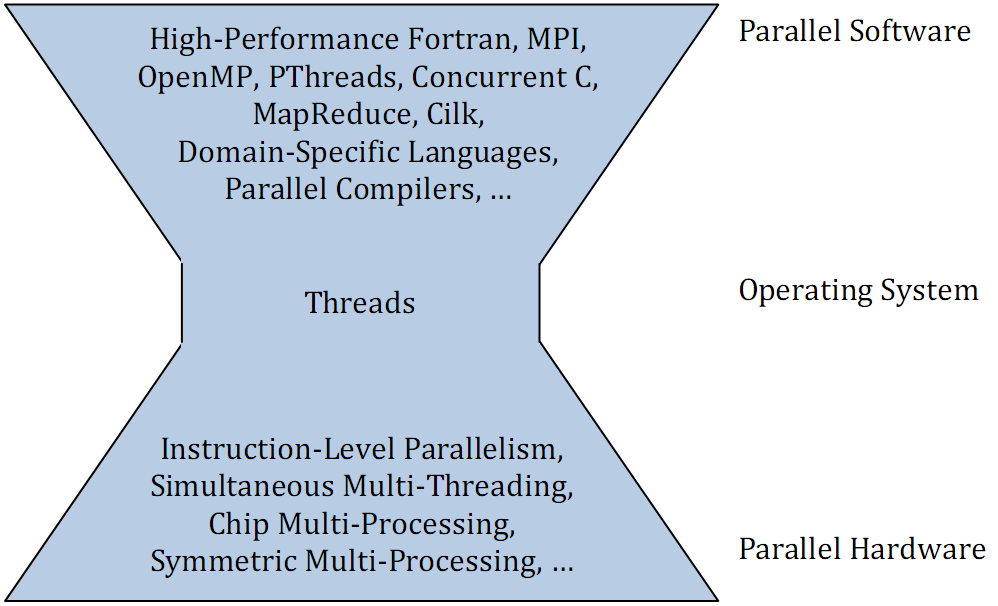}
\caption{The hourglass of parallelization support}
\label{fig:hourglass}
\end{figure}

The next level of software parallelism support is the operating system, which currently acts as common glue between hardware parallelism and application parallelism (see Figure \ref{fig:hourglass}). All modern systems support preemptive scheduling of parallel activities, commonly named as \emph{threads} \cite{distributed_os}. This does not free the application from actually creating and maintaining these parallel activities, but it allows an abstract usage regardless of the particular underlying hardware. Multithreading is still the major parallelization paradigm, support by standardized libraries and virtual runtime environments. While early multi-threading libraries performed the task scheduling by themself, it is meanwhile common the map application-level parallelism directly to operating system threads. 

Running more and more software in such a multi-threaded fashion leads to different kind of processing workloads, all with different memory access patterns. One example are virtualization products, were a highly varying load from the guest operating system is mapped to one operating system process in the host system. Iyer et al. \cite{qosmemhierarchy} predict this to be a major problem for optimal resource utilization. As one possible solution, they propose the priorization of some threads on hardware level. This allows an optimized usage of cache space and memory bandwidth, but demands according hints from the operating system. 

On application level, there is a variety of possibilities to express concurrent activities. The can be roughly categorized in dedicated parallel programming languages and sequential language extensions. 

For the large base of sequential programs written in C- or Java-style languages, it is common to use libraries or abstractions for threads. One popular example is the OpenMP \cite{openmp} language extension, other examples are Cilk \cite{cilk} and Concurrent C \cite{concurrentc}. 

More radical approaches critizise threads as wrong kind of abstraction, and rely instead on approaches such as pure data-parallel programming \cite{hpf, mcuda}, functional programming \cite{parfuncprog} or reactive programming \cite{gsem04}. A third class of researchers elaborates on the notion of \emph{domain-specific languages}, in order to encapsulate the parallelization in high-level language constructs. Popular examples in this category are SQL, MapReduce \cite{mapreduce}, or Simulink.  It remains an open research question which of these approaches is most suitable for the programming of scalable commodity applications based on thousands of execution units. In fact, most implementations still end up in mapping the activities to operating system threads. Alternative solutions mostly demand a virtual runtime environment for the context switching \cite{stacklesspython}.

In general, application developers have to decide upon the degree of control they want to have, starting from low-level locking primitives up to implicit parallelism in functional languages. The low-level approach is still the dominating idea in high-performance parallel computing, since it is the only way to get maximum performance. The more abstract solutions are about to become more popular for the broader mass of developers. The choice influences also the design of according debugging facilities -- low-level parallelisation support demands an according support for investigating problematic parallel activities \cite{choi98deterministic}.

From a theoretical perspective, the best approach for an parallel application is the usage of a dedicated parallel programming language. Many past initiatives in high-performance computing tried to introduce according solutions, in order to get the maximum benefit from the given compute power. This leaded to a long history of libraries and parallel programming languages, all trying to abstract basic concurrency and synchronization issues for the developer. A few of them remained successful \cite{hpf, mpi2,jones:88}, most of them are forgotten. Future research for multi-core enabled software needs to remember the successful and - especially - the failed attempts for the abstraction of parallel processing issues, in order to learn from the past. A number of language proposals are still promising \cite{erlang,esterel}, and now need to be considered for future programming concepts. At the moment, there is still no agreement about the feasibility of such languages for average industrial developers \cite{sutter:concurrency-revolution:acmqueue:2005}.

Beside the way of expressing parallel activities, it is also still relevant to work on scalable algorithms. Latest research on \emph{lock-free} data structures \cite{lockfree} and software transactional memory \cite{larus:tmbook:mcp:2006} shows that there is still a lot of potential in scalable algorithm and data structure design. The high-performance computing field has a long tradition in this area, such as with linear algebra computation, spectral methods, matrix calculations, atmosphere modeling based on partial differential equations, VLSI floorplan optimization or graph traversal \cite{fosterparprog}. It is therefore necessary to consider this huge body of knowledge for the upcoming era of parallel computers. 

\section{Summary}

Multi-core architectures are meanwhile the state-of-the-art in IT hardware. Computers of the future will contain thousands of execution units per chip, either homogeneus or specialized for particular purposes. The parallelization of application workload is about to become the most relevant strategy for speedup and scaleup in every kind of application. Since single thread performance is no longer given for free, it is inavoidable to think and program in a parallel fashion. 

'The free lunch is over.' \cite{freelunch}.

The future research for multi-core enabled applications must be driven from two sides  - a better expression of hardware, and a better design of software. The increasing variety of CMP processor architectures must be abstracted in better models, in order to allow an appropriate software design. Software on the other side must become analyzable and designable according to the parallel workload it produces. This includes the formulation of design patterns \cite{parpatterns} and algorithms prepared for parallel execution. Some applications will be scalable by default -- 3D graphics, scientific computing or high-throughput server computing. The interesting challenges are all the other ones. 

\bibliographystyle{plain}
\bibliography{../literature/literature,../literature/dcl}

\end{document}